\documentclass[aps,prb,twocolumn,floatfix,floats,showpacs,superscriptaddress]{revtex4}
\usepackage{graphicx,latexsym}
\usepackage{amsmath}
\begin{document}

\title{A state with negative binding energy induced by\\
       coherent transport in a quantum wire}

\author{Vidar Gudmundsson}
\affiliation{Science Institute, University of Iceland,
        Dunhaga 3, IS-107 Reykjavik, Iceland}
\author{Chi-Shung Tang}
\affiliation{Physics Division, National Center for Theoretical
        Sciences, P.O.\ Box 2-131, Hsinchu 30013, Taiwan}
\author{Andrei Manolescu}
\affiliation{Science Institute, University of Iceland,
             Dunhaga 3, IS-107 Reykjavik, Iceland}

%

\begin{abstract}
In a two-dimensional quantum wire in a perpendicular magnetic field
with a smooth embedded repulsive scattering potential we find 
in the multimode conductance resonances caused by bound states with 
negative binding energy. These resonances are the counterexamples to 
well known dips in the conductance and evanescent states caused by 
quasi-bound states of attractive scattering centers in the wire.
\end{abstract}


\maketitle
%
%

%
%
Coherent quantum transport of electrons through quantum wires has been measured and
modeled by many groups during the last one and half decade.
The conductance quantization and deviations from it have been a focus
of many researchers. Early on, it was discovered that an attractive 
scatterer can block a conduction channel totally for a narrow range of
energy just at the end of a conduction plateau or 
step.\cite{Bagwell90:10354,Gurvitz93:10578}  
This sharp dip structure was analyzed to be caused by a total reflection
of the incoming electron wave in one channel due to a quasi-bound state
originating from the evanescent mode in the next higher energy subband.  

Without an external constant magnetic field nothing corresponding to
this blocking can occur in a quantum wire with an embedded repulsive
scatterer as it has no quasi-bound states in contrast to the attractive
scatterer. In an external magnetic field this argument does not hold 
as can be inferred by a publication of Laughlin.\cite{Laughlin83:3383}

In this publication we will show that, indeed, the transport properties of
a quantum wire with an embedded repulsive scatterer in a constant external
magnetic field can exhibit signs of quasi-bound states. 
We will demonstrate that these states can reveal their presence both by
resonance transmission peaks and by dips in the conductance indicating
resonant backscattering processes.

%
%
We consider a multi-mode transport of electrons along the $x$-direction  
through a two-dimensional quantum wire defined by a parabolic confinement 
in the $y$-direction with the characteristic energy $E_0 = \hbar\Omega_0$.
The electrons incident from the left ($x\rightarrow -\infty$) impinge on a
smooth Gaussian scattering potential $V_{sc}=V_0 \exp{(-\beta r^2 )}$
Together the magnetic field ${\bf B}=B{\hat{\bf z}}$ and the parabolic
confinement define a natural
length scale $a_w=\sqrt{\hbar/(m^*)\Omega_w}$, where 
$\Omega_w=\sqrt{\omega_c^2+\Omega_0^2}$, with the cyclotron frequency
$\omega_c=eB/(m^*c)$, is the natural frequency of the quantum wire in a
magnetic field. The length $a_w$ can be considered as an effective
magnetic length in the wire system.
A mixed momentum-coordinate presentation of the wave
functions $\Psi_E(p,y)=\int dx\:\psi_E(x,y)e^{-ipx}$ and expansion in
channel modes $\Psi_E(p,y)=\sum_n\varphi_n(p)\phi_n(p,y)$
in terms of the eigenfunctions for the pure parabolically confined wire in
magnetic field leads to a coupled set of Lippman-Schwinger integral equations
in momentum space. These equations are then transformed
into integral equations for the $T$-matrix in order
to facilitate numerical evaluation.\cite{Gurvitz95:7123,Gudmundsson05:BT} 
The conductance is evaluated according to the Landauer-B{\"u}ttiker 
formalism together with the scattering wave functions
from the $T$-matrix.\cite{Gudmundsson05:BT}

%
%
We consider a broad parabolic quantum wire with confinement energy
$\hbar\Omega_0 = 1$ meV. At a vanishing magnetic field $B = 0$ this energy
corresponds to $a_w = 33.7$ nm. We select a fairly narrow but smooth 
scatterer in the center of the wire with $\beta = 1\times 10^{-2}$ nm$^{-2}$,
equivalent to $\beta a_w^2 = 11.4$ at $B = 0$, or in other words the 
scattering potential has reached $e^{-1}$ at $r\approx 10$ nm, close to the
value of the effective Bohr radius in GaAs $a_0^* = 9.79$ nm.  
In the numerical calculations we have included at least 13 channel modes
in the wire and use an unevenly spaced grid in momentum space to
apply a repeated 4-point Gaussian integration to the integral equations
for the T-matrix. Numerical accuracy is assured through comparison to
solutions obtained with a larger basis set. 

\begin{figure}[tbhp!]
      \includegraphics[width=0.48\textwidth,angle=0]{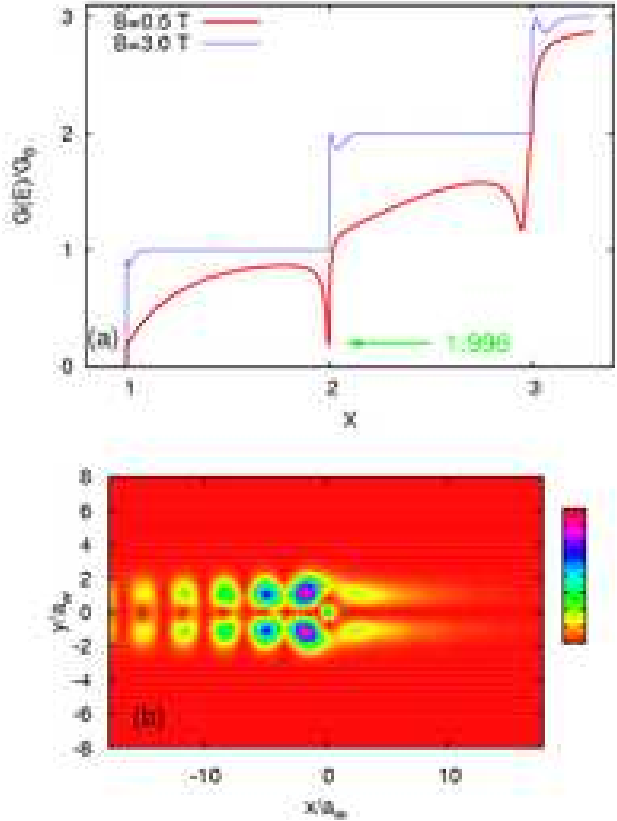}
      \caption{(Color online) (a) The conductance in units of $G_0=2e^2/h$
               of a broad wire with one embedded narrow 
               potential well at the center $x=0$,
               as function of $X=E/(\hbar\Omega_w)+1/2$.  
               (b) The probability density of the scattering state
               corresponding to an incident state
               with $X=1.996$ and $n=1$.
               $E_0=\hbar\Omega_0=1.0$ meV, $V_0=-8$ meV,
               $\beta = 10^{-2}$ nm$^{-2}$, and for GaAs $a_0=9.79$ nm.
               $a_w=29.34$ nm at $B=0.5$ T, and $a_w=14.68$ nm at $B=3.0$ T.}
\label{Fig_1}
\end{figure}

In order to understand better the results for a wire with a repulsive
scatterer we first show in Fig.\ \ref{Fig_1} results for a wire with
an {\em attractive} scatterer. 
To compare conductance curves for different values of $B$ we show them
as functions of $X = E/(\hbar\Omega_w)+1/2$.
The integer part of the parameter $X$ counts how many channels in the
wire are open for transport for an incoming electron with 
propagating energy $E$.  
At low magnetic field, $B = 0.5$ T, dips are seen in the conductance curve
just before $X$ assumes integer values. These well known dips can in
a perturbative picture be explained as being caused by a backscattering
of the incoming electron in subband $n$ by an evanescent state in subband 
$n+1$ that the scattering potential has lowered into the gap between
the subbands just below the $n+1$ subband. 
The probability density of the electronic state in the
first subband with $X=1.996$ is shown in Fig.\ref{Fig_1}(b).
On the incident side, the left side, we see the interference pattern of
the incoming and outgoing $n=1$ state, and on the right side there is only
the evanescent probability. 

For a higher magnetic field, $B = 3$ T, the Lorentz force corresponding to
the kinetic energy necessary to reach the evanescent state is large enough
to press the electron into one side of the wire far enough from its center
suppressing the overlap between the incoming and the evanescent state. 
As a result there is no dip found, and the only clear deviation from
perfect conductance is in the beginning of each conductance step where
the Lorentz force is small enough to still allow for an encounter between
the electron and the scattering potential.      

This together with the curious fact noted by Laughlin\cite{Laughlin83:3383}
that two electron restricted to a plane perpendicular to a constant magnetic
field can form a bound state with negative binding energy leads us to the
following train of thought: If two electrons form a bound state, so can
also one do around a smooth potential hill. The potential hill in the middle
of a quantum wire will lift at least one state higher into the band where
it originated. This quasi-bound state just above the band minimum might 
influence the scattering at high magnetic field just in the beginning
of a new conductance step. In a flat two-dimensional system the state
would be a true bound state, but in a quantum wire there is always an
equipotential line along the wire edge with the same energy as the 
bound state guaranteeing at least a vanishingly small overlap between
the edge and the bound state.  
   
\begin{figure}[tbhp!]
      \includegraphics[width=0.48\textwidth,angle=0]{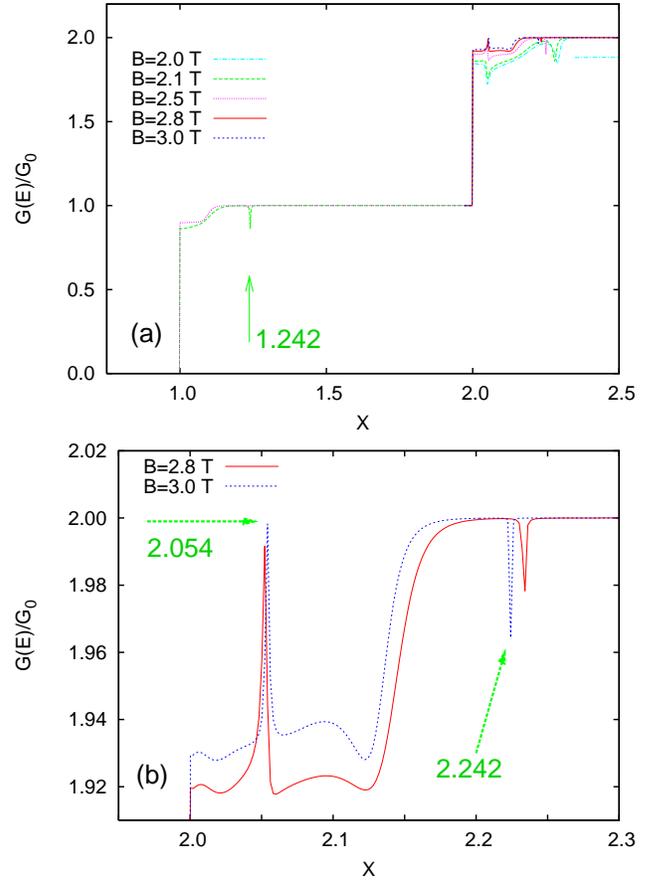}
      \caption{(Color online) The conductance in units of $G_0=2e^2/h$
               of a broad wire with one embedded narrow 
               potential hill at the center $x=0$,
               as function of $X=E/(\hbar\Omega_w)+1/2$ (a), and
               the same repeated for a narrow range of $X$. 
               $E_0=\hbar\Omega_0=1.0$ meV, $V_0=+8$ meV,
               $\beta = 10^{-2}$ nm$^{-2}$.}
\label{Fig_2}
\end{figure}

\begin{figure}[tbhp!]
      \includegraphics[width=0.48\textwidth,angle=0]{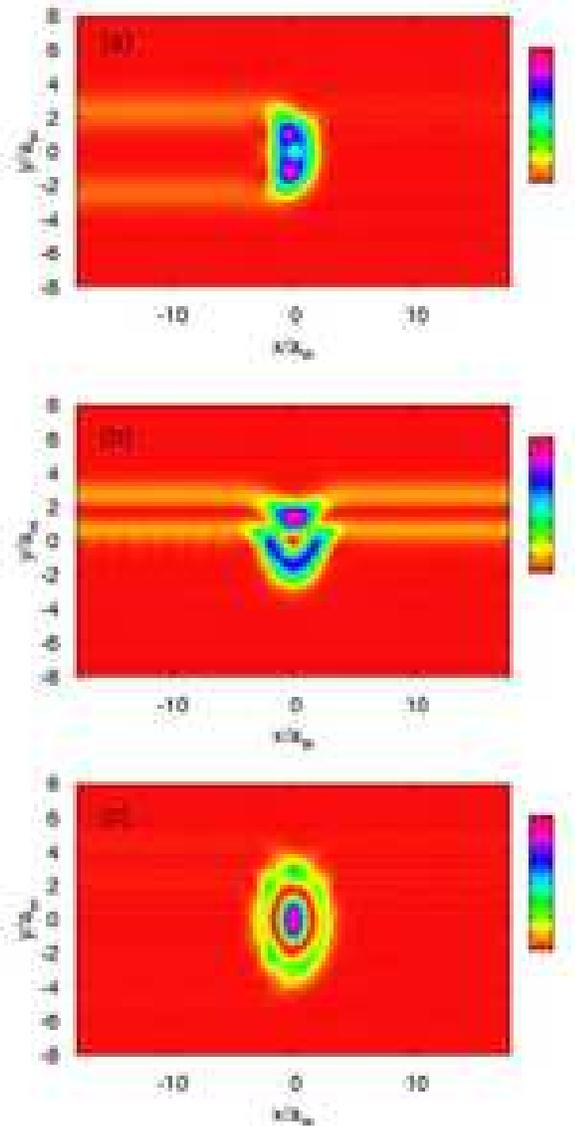}
      \caption{(Color online) Corresponding to special values
               of $X$ in Fig.\ \ref{Fig_2}  
               the probability density of the scattering state
               corresponding to an incident state with $X=1.242$, 
               $n=1$, and $B=2.0$ T (a), $X=2.054$, $n=2$, and 
               $B=3.0$ T (b), and $X=2.242$, $n=2$, and 
               $B=3.0$ T (c).
               $E_0=\hbar\Omega_0=1.0$ meV, $V_0=+8$ meV,
               $\beta = 10^{-2}$ nm$^{-2}$.}
\label{Fig_3}
\end{figure}

The parameters chosen earlier in this paper for the wire and the
scattering potential proved fruitful in the search for a bound
state of an electron around a hill. In Fig.\ \ref{Fig_2}(a) we
show the conductance of a wire with an embedded potential hill,
a repulsive scatterer. A fine structure is visible at the beginning
of each conductance step. In Fig.\ \ref{Fig_2}(b) we focus in on the
beginning of the second step for two close values of $B$.
The arrows point at a transmission resonance at $X = 2.054$ and
a dip at $X = 2.242$ for $B = 3.0$ T.  We shall also analyze a
dip at the beginning of the first step at $X = 1.242$ for 
an electron entering the wire in the $n = 1$ channel mode
at $B = 2.0$ T. The corresponding probability densities for
the electronic states are seen in Fig.\ \ref{Fig_3}.

The kinetic energy of incoming electrons close to the transmission 
peak at $X = 2.054$ is low leading to a small Lorentz force acting on them.
The electrons thus encounter the scattering potential and there is
some backscattering, though small. The probability density in 
Fig.\ \ref{Fig_3}(b) and the corresponding transmission
peak in Fig.\ \ref{Fig_2} shows us that at $X = 2.054$ the electrons  
come into resonance with a quasi-bound state with binding energy of
approximately $-0.29$ meV. This low energy and the potential curvature of the
wire give the probability density a shape that reflects both the 
structure of the resonant edge state and the circular scattering potential.   
The double hump structure of the incoming and the transmitted wave 
are signs of the $n = 2$ mode or channel, the $n = 1$ channel is further away
from the center of the wire as is discussed below.
 
At higher kinetic energy for the incoming electrons the Lorentz force is 
large enough to allow most of them to bypass the potential resulting in
almost perfect conductance, except for the dip at $X = 2.242$ where the
electrons come into resonance with a higher lying quasi-bound state
that causes some of them to be backscattered. The probability density of
this state is displayed in Fig.\ \ref{Fig_3}(c). It is long lived judging
from the narrowness of the dip and the strength of the probability density
that subdues almost the incoming and the reflected wave on the color scale
used here. The binding energy
of this state is $-1.18$ meV and the smaller coupling to the edge states
makes its symmetry much closer to the circular symmetry of the scatterer.
The double ring structure of this state suggests that there are more
quasi-bound states with simpler structure and larger negative binding
energy but their still smaller coupling to the edge states makes them
invisible to our calculation. 

A simple estimate of the binding energy of the scattering potential
neglecting the confinement of the wire and using the semiclassical 
quantization condition of Bohr places the value of $-1.18$ meV in between
values obtained by assuming the quantum numbers 2 and 2+1/2. 
This is in accordance with the double hump 
structure we see in the probability density in Fig.\ \ref{Fig_3}(c).  
 
It should be stated here that the probability densities for the 
$n = 1$ states corresponding to the $n = 2$ states in 
Fig.\ \ref{Fig_3}(b) and (c) at $B = 3$ T are straight edge states
with their maxima at $y \approx 8a_w$. So, we are observing a 
scatterer with radius approximately $0.8a_w$, the bound state in
Fig.\ \ref{Fig_3}(b) has a radius of approximately $3a_w$, and 
the distance from the bound state to the edge is approximately $5a_w$ 
($a_w = 14.7$ nm here). So, indeed, our scatterer is narrow but smooth
leading to quasi-bound states of simple structure.  

In Fig.\ \ref{Fig_3}(a) is the probability density corresponding to 
to the dip at $X=1.242$ in the first subband. We can faintly see the
incoming and the reflected waves. Here $B = 2.0$ T and due to the greater
magnetic length or $a_w$ than for $B = 3.0$ T the symmetry of the quasi-bound state
is influenced by the edge states. The resonance dip for this structure is
deeper at $B = 2.1$ T (see Fig.\ \ref{Fig_2}(a)) but the faint probability for
the incoming and the outgoing wave at $B = 2.0$ T shows us the 
effective width of the wire here for this $n=1$ state.   

\begin{figure}[tbhp!]
      \includegraphics[width=0.48\textwidth,angle=0]{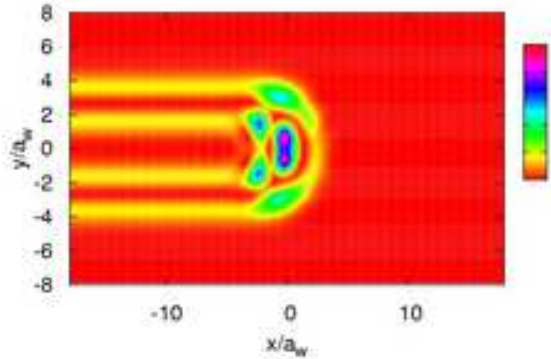}
      \caption{(Color online) 
               The probability density of the scattering state
               corresponding to an incident state with $X=2.29$, 
               $n=2$, and $B=2.0$ T corresponding to a broad
               resonance in Fig.\ \ref{Fig_2}(a) 
               $E_0=\hbar\Omega_0=1.0$ meV, $V_0=+8$ meV,
               $\beta = 10^{-2}$ nm$^{-2}$.}
\label{Fig_4}
\end{figure}

The resonance at $B = 3$ T in Fig.\ \ref{Fig_2}(a) corresponding to the state in 
Fig.\ \ref{Fig_3}(c) is narrow reflecting a long lived state. At a lower
magnetic field $B = 2$ T the corresponding resonance is broad indicating a 
relatively short lived resonance state producing it. Exactly this can be  
verified by looking at Fig.\ \ref{Fig_4},
where the incoming and outgoing waves in the $n=2$ channel are clearly seen
and one can infer from the structure of the probability density that 
semiclassically speaking the
electrons undergoe few reflections before being returned.

%
%
We have demonstrated that quasi-bound states of a repulsive potential hill
in a quantum wire in an external magnetic field can leave their fingerprints 
on the conductance of the system for suitably selected parameters.
A small coupling to the edge states gives a long lived 
quasi-bound state with the symmetry of the scatterer, but the small
coupling may hinder the state in affecting the conductance of the system. 
The narrow but smooth potential hill in an external magnetic field behaves like a 
``quantum peg'' to which the electrons can be hooked for some time
in contrast to the more familiar picture of the quasi-bound state of
a quantum dot where one imagines the electron rattling for a short time
in a bowl-like structure, the dot.

The use of a smooth scattering potential with feature lengths comparable 
to the effective Bohr radius in the system has allowed us to make visible
the effects of fairly simple ``low-lying'' quasi-bound states at intermediate
strengths of magnetic field $B\approx 3$ T. Clearly the effects of these
quasi-bound states on the conductance are smaller than the effects 
of the corresponding states of an attractive scatterer so they are probably 
not easy to detect experimentally. 

Here we have focused our view on the bound state of a smooth repulsive 
scatterer in a quantum wire in a perpendicular magnetic field as a
complimentary system to the wire with an embedded attractive scatterer,
and have compared the effects of these two scatterers on the conductance
of the system. A similar, system has been considered analytically in the 
extreme quantum limit where a one band limit could be applied by Jain and 
Kivelson in order to explain the break down of the quantum Hall effect in narrow
constrictions.\cite{Jain88:1542} Takagaki and Ferry used a hard wall
version of both a quantum strip and a scatterer to explore with a tight-binding 
formalism how a circulating edge channel around an antidot leads to an 
Aharonov-Bohm-type oscillation of the conductance as the magnetic field 
normal to the strip is varied.\cite{Takagaki93:8152} They assume the
diameter of the antidot is large compared to the characteristic wavelength
of the electrons and focus their attention on the magnetic coupling of the
the electron waves through the constriction on each side of the antidot.
Corresponding experimental system with two large antidots has been 
investigated by Gould et al.\cite{Gould95:11213} 
  
We expect that a larger hard wall scatterer may also produce quasi-bound 
states (with edge state character around the scatterer) 
that could influence the conductance of the system, but we believe that 
those states might produce effects in the
conductance that are harder to find due to their very low binding energy and
closeness to the beginning of a conductance plateau.


%
%
\begin{acknowledgments}
      The research was partly funded by the Research
      and Instruments Funds of the Icelandic State,
      the Research Fund of the University of Iceland, and the
      National Science Council of Taiwan.
      C.S.T. acknowledges the computational facility supported by the
      National Center for High-performance Computing of Taiwan.
\end{acknowledgments}

%
%
\bibliographystyle{apsrev}

%
%
%
\end{document}